# Transcranial magnetic stimulation of visual-motion area V5/MT modulates sensory thalamus responses during visual speech recognition

Lisa Jeschke[1], Christa Müller-Axt[1], Alejandro Tabas[234], Begoña Díaz[5], and Katharina von Kriegstein[1]

[1] *Chair of Cognitive and Clinical Neuroscience, Faculty of Psychology, Dresden University of Technology, Dresden, Germany*
[2] *Basque Center on Cognition, Brain and Language, San Sebastian, Spain*
[3] *Max Planck Institute for Human Cognitive and Brain Sciences, Leipzig, Germany*
[4] *Ikerbasque, Basque Foundation for Science, Bilbao, Spain*
[5] *Universitat Internacional de Catalunya, Faculty of Medicine and Health Sciences, Barcelona, Spain*

## Abstract

Responses in the sensory thalamic nuclei are modulated by perceptual tasks. Whether such response modulations rely on feedback from cerebral cortex in humans is unknown. Here, we addressed this question in the context of visual speech recognition: the visual sensory thalamus, i.e. the lateral geniculate nucleus (LGN), has differential BOLD-responses to visual speech than non-speech control tasks. We tested whether such response modulation relies on the function of the visual association cortex, specifically visual-motion area V5/MT.

We applied inhibitory transcranial magnetic stimulation (TMS) over bilateral visual-motion sensitive areas V5/MT on 26 healthy adults. Subsequently, participants performed a visual speech and a colour recognition task on identical muted videos of speaking faces during functional magnetic resonance imaging (fMRI). The LGN showed a significant signal change between the visual speech task and the colour task following Vertex stimulation as active control region. This modulation was significantly reduced following inhibitory V5/MT stimulation. V5/MT stimulation also reduced task-dependent functional connectivity between V5/MT and the LGN.

These results identify corticothalamic feedback as integral mechanism in visual processing. In particular, the visual association cortex has a causal role in modulating LGN responses during speech recognition.

## Introduction

Speech recognition in human face-to-face communication involves the processing of not only auditory, but also visual speech signals [1,2]. Visual speech signals have a central role in communication as they facilitate the recognition of auditory speech, especially in noisy environments [2,3] or for hearing impairments [4].

Speech recognition modulates blood oxygenation-level dependent (BOLD) responses of the visual and auditory sensory thalami [5–7]. For instance, the signal change of the visual sensory thalamus, i.e. the lateral geniculate nucleus (LGN), is larger when participants recognise visual speech compared to the speaker's identity, despite viewing identical muted talking faces [7]. In the following, we will call such modulation 'task-dependent modulation'. It is an unresolved question where such task-dependent modulation of thalamic nuclei originates. A plausible account is that it stems from feedback of specialised cerebral cortical regions [8,9]. Research using cell-recordings and animals demonstrated that the LGN receives feedback from the striate and extrastriate cortex [10–12]. For instance, inactivation of the middle temporal area V5/MT changes LGN responses in non-human primates [13]. To our knowledge, however, there is no direct evidence for such cortical feedback in humans, particularly during active cognitive tasks. Hence, the aim of the present study was to test whether task-dependent

LGN modulation for visual speech stems from extrastriate cortices, i.e. visual-motion sensitive area V5/MT. Visual speech recognition relies on the processing of dynamic and configural facial features, associated with respective dorsal and ventral visual regions, including area V5/MT [14]. Studies using non-invasive transcranial magnetic stimulation (TMS) demonstrated a causal influence of V5/MT on the recognition of non-biological motion [15–17] as well as visual speech [18]. Moreover, direct connections between V5/MT and the thalamus have been demonstrated in non-human primates and humans [19–21] and have been implied by human TMS and blindsight studies [16,22–24]. V5/MT might thus be a candidate region for modulating the LGN during visual speech recognition.

We applied offline inhibitory TMS over bilateral V5/MT, followed by functional MRI (fMRI) while participants performed two tasks with identical stimuli - muted videos of speakers talking. Depending on task instructions, participants carried out a visual speech (Speech Task) or a facial colour recognition task (Colour Task). We expected that, compared to active control stimulation, V5/MT stimulation would reduce task-dependent LGN modulation and task-dependent functional V5/MT-LGN connectivity.

## Methods

The study was approved by the local Ethics Committee (EK1701042019) and data were acquired at Technische Universität Dresden, Germany. Participants gave written informed consent and were reimbursed monetary or by student credit. The study was preregistered [41].

### Participants

Twenty-six native German speaking, healthy, right-handed adults (20 female, 6 male) between 18 and 38 years of age (M = 25.22, SD = 4.81) were included in the final sample. All participants were screened for conditions associated with altered visual speech recognition or colour vision (Supplementary Information).

### Procedure

The study comprised three sessions (Figure 1): Session 1 included MR image acquisition for localizing V5/MT and the LGN. Sessions 2 and 3 started with an experimental task outside the MRI machine (Baseline). We then applied TMS over bilateral V5/MT or the vertex as active control site. Subsequently, participants performed the same experimental tasks during fMRI (Speech and Colour Task). The order of stimulation sessions was counterbalanced across participants.

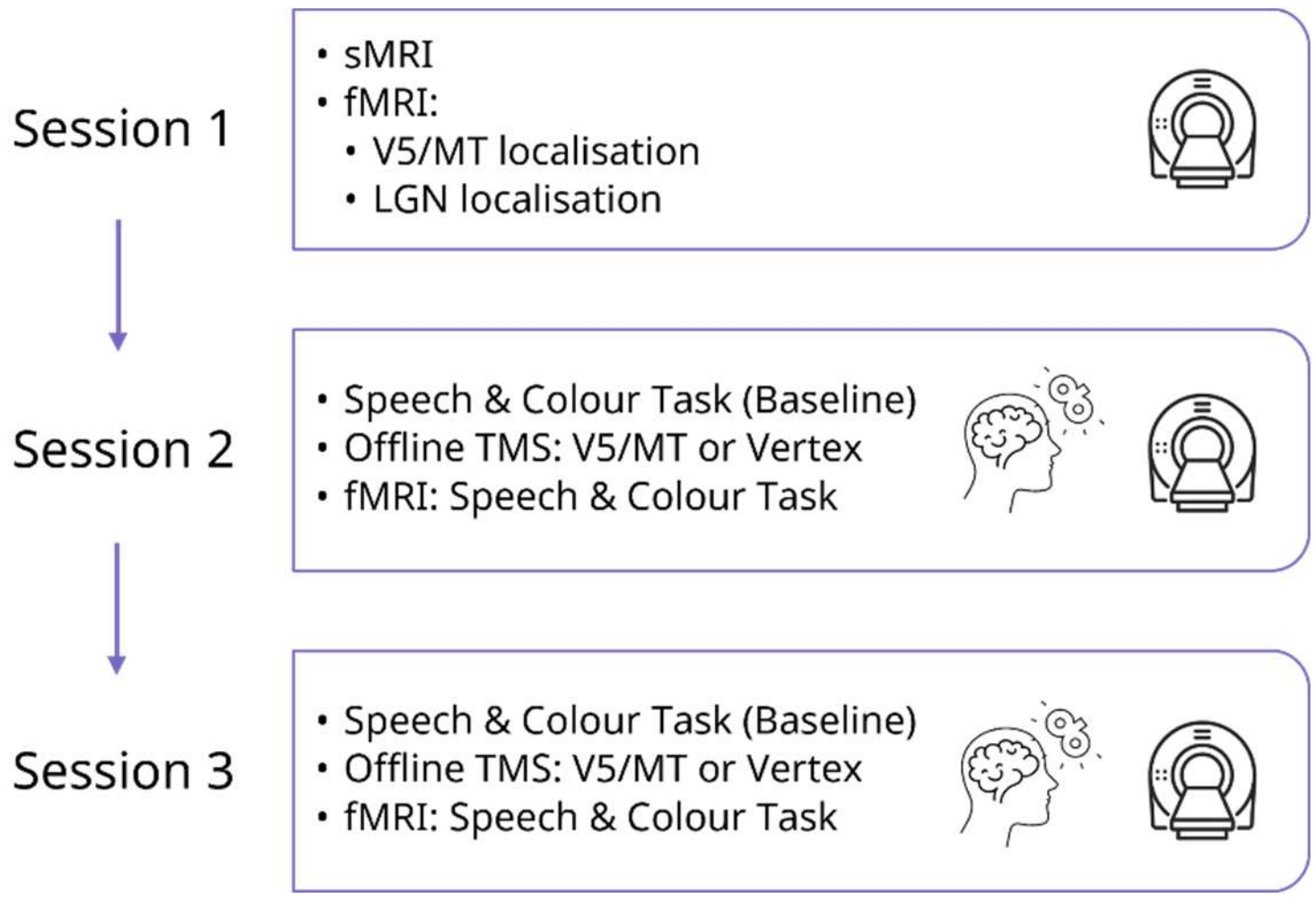

*Figure 1.* Summary of the experimental sessions. All participants underwent structural (sMRI) and functional MRI (fMRI) in their first session. In Sessions 2 and 3, we first measured participants' baseline performance in the Speech and Colour Task. This was followed by the application of offline transcranial magnetic stimulation (TMS) over bilateral area V5/MT or the Vertex region in a within-subject design. Subsequently, participants performed the Speech and Colour Task during fMRI. Sessions 2 and 3 were separated by a mean of 12.26 days (SD = 7.96, range = 3-28). At the end of Session 3, participants completed a questionnaire assessing perceived differences between TMS conditions, side effects, and task strategies (Supplementary Information).

## Experimental Design - Speech and Colour Task

**Stimuli.** Stimuli were muted videos of one of three male native German speakers presented against a dark-grey oval with a black background (Figure 2; example: https://osf.io/mbfvp/). Speakers produced short vowel-consonant-vowel syllables (36 in total). All vowels (/a/, /e/, /u/) and consonants (/f/, /p/, /n/, /s/) corresponded to discriminable German viseme classes - a viseme being defined as a set of phonemes identical to the interlocutor [25,26]. Videos were edited (Video Editor Pro v6.8.2.341; Flash-Integro LLC) to colourise the speakers using six semi-transparent colours at equal intervals along a yellow-pink spectrum (RGB: 250/235/190, 250/220/190, 250/205/190, 250/190/190, 250/190/205, 250/190/220), preserving natural skin tone appearance and using a dynamic mask that prevented colour overlap with the background.

**Task.** Participants performed two 1-back tasks: a Speech Task and a Colour Task (Figure 2). Tasks were presented in blocks (8 videos/block) that began with an instruction screen ("SILBE" for Speech Task or "FARBE" for Colour Task). For the Speech Task, participants were asked to respond as fast and correct as possible whether the uttered syllable was identical or different to the syllable in the previous video. Analogously, in the Colour Task, they decided whether the facial colour was identical or different to the one in the preceding video, independent of the spoken syllable. The task blocks were presented randomly. Responses were made via button press, counterbalanced across participants. Tasks were aligned in difficulty. Stimuli and within-block randomisation were identical across tasks (Supplementary Information). The match rate was jittered 43% - 57% per block, resulting in an overall 50% chance-level accuracy. The experiment was programmed in Presentation (Neurobehavioural Systems, Inc., USA). Outside the MRI device, participants completed two practice blocks followed by 12 baseline blocks before TMS, and 54 experimental blocks (24 min in total) during fMRI. To ensure comparable eye movement patterns across tasks, eye-tracking data were collected from three different participants prior to the main data acquisition. No significant interaction of task and face area was observed for the duration and number of fixations (Supplementary Information).

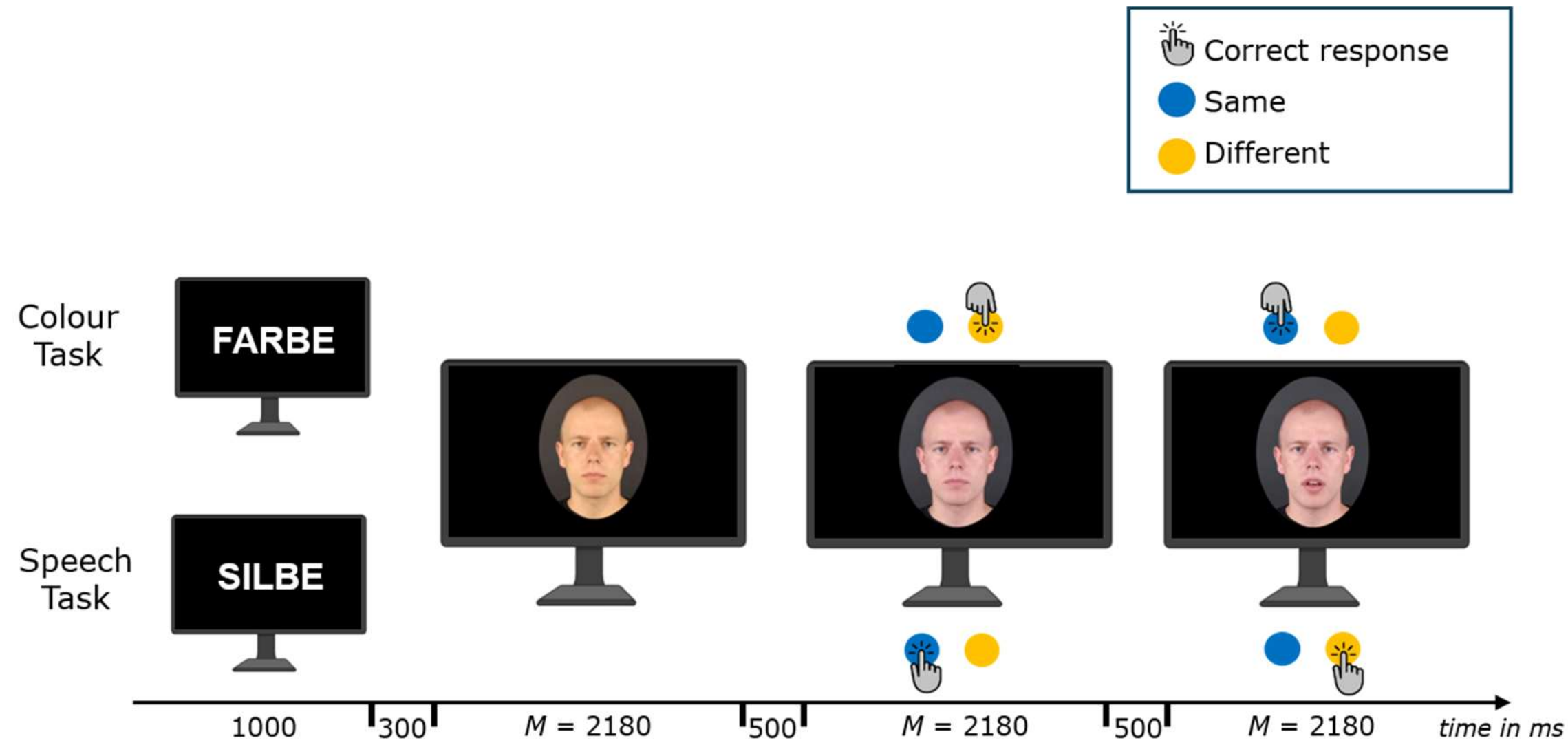

*Figure 2.* Example trials for the Speech and Colour Tasks. Participants performed two 1-back tasks with identical stimuli: in the Speech Task ("SILBE"; English: "syllable"), participants responded whether the uttered syllable in the muted video matched or differed from the one in the previous video. In the Colour Task ("FARBE"; English: "colour"), participants indicated whether the facial colour of the speaker was the same as or different from the immediately preceding stimulus. Videos had an average duration of 2180 ms and were separated by 500 ms inter-trial intervals (average block duration 20.14 s).

### Transcranial Magnetic Stimulation

Offline TMS was administered in Sessions 2 and 3. At the beginning of Session 2, active motor thresholds (AMT) were determined to set individual stimulation intensity (Supplementary Information). Coil positioning was guided by stereotactic neuronavigation (BrainSight, Rogue Research, Canada) using individual coordinates for V5/MT, vertex, and primary motor cortex (Supplementary Information). We chose an active stimulation as control site to rule out effects that would be caused by general stimulation compared to sham stimulation, based on established inhibitory Vertex protocols [27–29].

**TMS parameters.** Continuous theta-burst stimulation (cTBS) was delivered using a MagPro X100 stimulator (MagVenture, Germany) with a figure-of-eight coil (MCF-B65, 75 mm). Stimulation intensity was set to 100% of the AMT. Each cTBS protocol consisted of 200 frames (60 ms) separated by 140 ms intervals, resulting in a total duration of 40 s. Each frame contained three pulses at 15 ms intervals (600 pulses total). Immediately after stimulation, participants entered the MRI scanner and performed the Speech and Colour Tasks.

### MR Image Acquisition

Structural and functional images were acquired on a 3T MAGNETOM Prisma (Siemens Healthineers AG, Forchheim, Germany) with a 32-channel receive head coil. T1-weighted images were acquired with an MP-RAGE sequence (TR = 2.0 s; TE = 1.97 ms; flip angle = 8°; bandwidth = 238 Hz/Px) with a 1 $mm^3$ voxel size (256 sagittal slices). Functional images for the Speech and Colour Tasks were acquired using a gradient-echo EPI sequence (TR = 2.1 s, TE = 46 ms, flip angle = 79°, bandwidth = 1442 Hz/Px) at 1.5 $mm^3$ resolution across 24 axial slices positioned to cover V5/MT and the thalamus. Five full field-of-view volumes (84 slices) were acquired as intermediate images for preprocessing. Images for V5/MT localisation were acquired using a gradient-echo EPI sequence (TR = 869 ms, TE = 38 ms, flip angle = 58°, bandwidth = 1832 Hz/Px) with whole-brain coverage at 2.4 $mm^3$ resolution (60 sagittal slices). One field map (TE1/TE2 = 4.92/7.38 ms) was acquired per session.

**Functional localisation of V5/MT and LGN**. V5/MT was localised using a moving-versus-static paradigm [24], with random-dot kinematograms consisting of static or coherently inward/outward moving white dots on a black background (Supplementary Information). Stimuli for LGN localisation consisted of 16 blocks of high-contrast flickering radial checkerboards presented alternately in the left or right visual hemifield, while the opposite hemifield displayed a uniform grey background [31].

### Preprocessing and Analysis

**V5/MT localisation**. MRI data were preprocessed and analysed with SPM12 (Wellcome Trust Center for Neuroimaging, UK) in a MATLAB environment (R2019b; v9.6.0.1047502). Preprocessing included motion correction, realignment, unwarping, normalization to MNI space, and spatial smoothing (8 mm FWHM). Statistical parametric maps were generated using general linear models (GLMs), modelling the hemodynamic responses to moving and static conditions. For first-level analyses, the t-contrast *moving* > *static* was computed and individual stimulation coordinates were defined as the peak voxel ($p < 0.001$ uncorrected) within a V5/MT cluster derived from a previous study [18]. Coordinates were transformed into native space for neuronavigated TMS. Group-level analyses used a family-wise-error (FWE) corrected threshold of $p < 0.05$. Resulting clusters (left: 1824 $mm^3$; right: 1344 $mm^3$) were used as ROIs for blood oxygenation level dependent (BOLD) signal extraction.

**LGN localisation.** Functional LGN localisation did not reliably identify bilateral LGN activation for all participants (i.e. widespread activation patterns or a lack of clear lateralized responses), hence a high-resolution probabilistic LGN atlas was used for all participants [31]. Atlas probability thresholds were adjusted separately for each hemisphere to obtain anatomically comparable volumes [32], resulting in thresholds of 40% (124.81 ± 10.81 $mm^3$) for the left LGN and 50% for the right LGN (128.27 ± 10.57 mm).

**Speech and Colour Task: fMRI data.** We segmented the structural images using FreeSurfer (version 7.3.2; [33]). Functional images were motion-corrected in SPM12, registered to structural images, and resampled using FreeSurfer [34]. Data were then smoothed with a 2 mm FWHM Gaussian kernel. We generated first-level statistical maps in SPM using a 2 × 2 repeated-measures design with the within-participant factors Task (Speech vs Colour) and Stimulation (V5/MT vs Vertex). Contrasts of interest were assessed using t-contrasts. For the group-level analysis, we transformed ROIs from MNI standard space into native space using ANTs [35]. Afterwards, we extracted the mean beta weight of the respective contrast within the individual ROI for every participant. Analyses were then conducted by means of a one-sample two-tailed t-test across the 1st level mean beta estimates in R (v3.6.3; [36]), with significance at $p < 0.05$ and Bonferroni-Holm corrections applied to post-hoc t-tests.

**Speech and Colour Task: Behavioural data.** Response times (RTs) were analysed as the primary behavioural measure, as TMS effects typically manifest in RTs rather than accuracy [37–39]. We defined RTs as the interval from video onset to response and excluded responses < 200 ms and incorrect trials. RTs were analysed both as absolute values during fMRI as well as after normalization to baseline performance (i.e. RT ratio). RT ratios were calculated as post/pre-stimulation RTs for each participant, task, and stimulation condition to account for practice effects and inter-individual variability. We analysed the dependent variables with linear mixed-effects models (LMM) in R (v3.6.3; [36]) that included fixed effects of Task and Stimulation. The random-effects structure of the LMMs was determined by removing random-effects terms accounting for the least variance until the fitted model converged (Supplementary Information). Effects from both models were Bonferroni-Holm corrected across models. We also conducted two-tailed ($p < 0.05$) correlation analyses between behavioural measures and mean beta estimates, Bonferroni-Holm corrected within each correlation hypothesis.

**Task-dependent functional connectivity.** We performed psychophysiological interaction (PPI) analyses [40] for both stimulation conditions. After creating GLMs on the single subject level, we extracted time series for left/right V5/MT and left/right LGN within the predefined ROIs for the contrast 'Speech Task - Colour Task'. The time series was deconvolved with the canonical hemodynamic response function (HRF) to approximate the underlying neural activity. Next, we created the psychophysiological interaction term by multiplying this deconvolved neural time series with the psychological vector coding the contrast of interest. The resulting interaction term was deconvolved with the HRF and entered into a new first-level GLM together with the seed time series and the task regressor as covariates. The contrast of interest tested whether the interaction term explained significant variance in the target ROI, indicating task-dependent connectivity. For every participant, we extracted the mean beta estimate within the corresponding ipsilateral ROI as target region, e.g., for the GLM created from left LGN as seed, we extracted the estimate for left V5/MT as ROI and vice versa. We then pooled the beta estimates for both hemispheres. Group-level analyses were conducted by means of a one-sample two-tailed t-test across the 1st level mean beta values. Effects were considered significant at $p < 0.05$, and Bonferroni-Holm corrected for the two seed regions.

## Hypotheses

All hypotheses on LGN BOLD responses, V5/MT BOLD responses, behavioural responses, and their correlations were preregistered [41]. Our main hypothesis was that V5/MT influences the task-dependent LGN modulation: we expected that, compared to Vertex stimulation, V5/MT stimulation would reduce the difference in LGN BOLD responses between the two tasks. In a follow-up investigation (not preregistered), we expected that this effect would be reflected in functional connectivity changes between V5/MT and the LGN. Second, we hypothesized that inhibitory TMS over V5/MT, relative to Vertex stimulation, would impair visual speech recognition, resulting in delayed RTs, with smaller effects in the Colour Task. We also expected a larger BOLD response of V5/MT for the Speech Task compared to the Colour Task. Additionally, we expected positive correlations between neural and behavioural outcomes for visual speech recognition.

## Results

**LGN responses.** Our main hypothesis was that cortical area V5/MT influences task-dependent LGN modulation. We expected that the difference in LGN response between tasks would be reduced following V5/MT compared to Vertex stimulation, tested by means of the interaction contrast "Vertex (Speech Task > Colour Task) > V5/MT (Speech Task > Colour Task)". Two-tailed t-tests across first-level beta estimates revealed a significant Task-Stimulation interaction ($t(24) = -2.254$, $p = 0.029$, $d = 0.31$; Figure 3): there was a significant Task effect under Vertex stimulation ($t(24) = 5.539$, $p < 0.001$ Bonferroni-Holm corrected for two post-hoc tests, $d = 0.78$). This effect became nonsignificant by V5/MT stimulation ($t(24) = 1.415$, $p = 0.653$ Bonferroni-Holm corrected for two post-hoc tests, $d = 0.22$). There was also a main effect of Task ($t(24) = -3790$, $p < 0.001$ Bonferroni-Holm corrected, $d = 0.74$). The Speech Task elicited a more negative signal change (mean beta weight = -0.399) than the Colour Task (mean beta weight = -0.177).

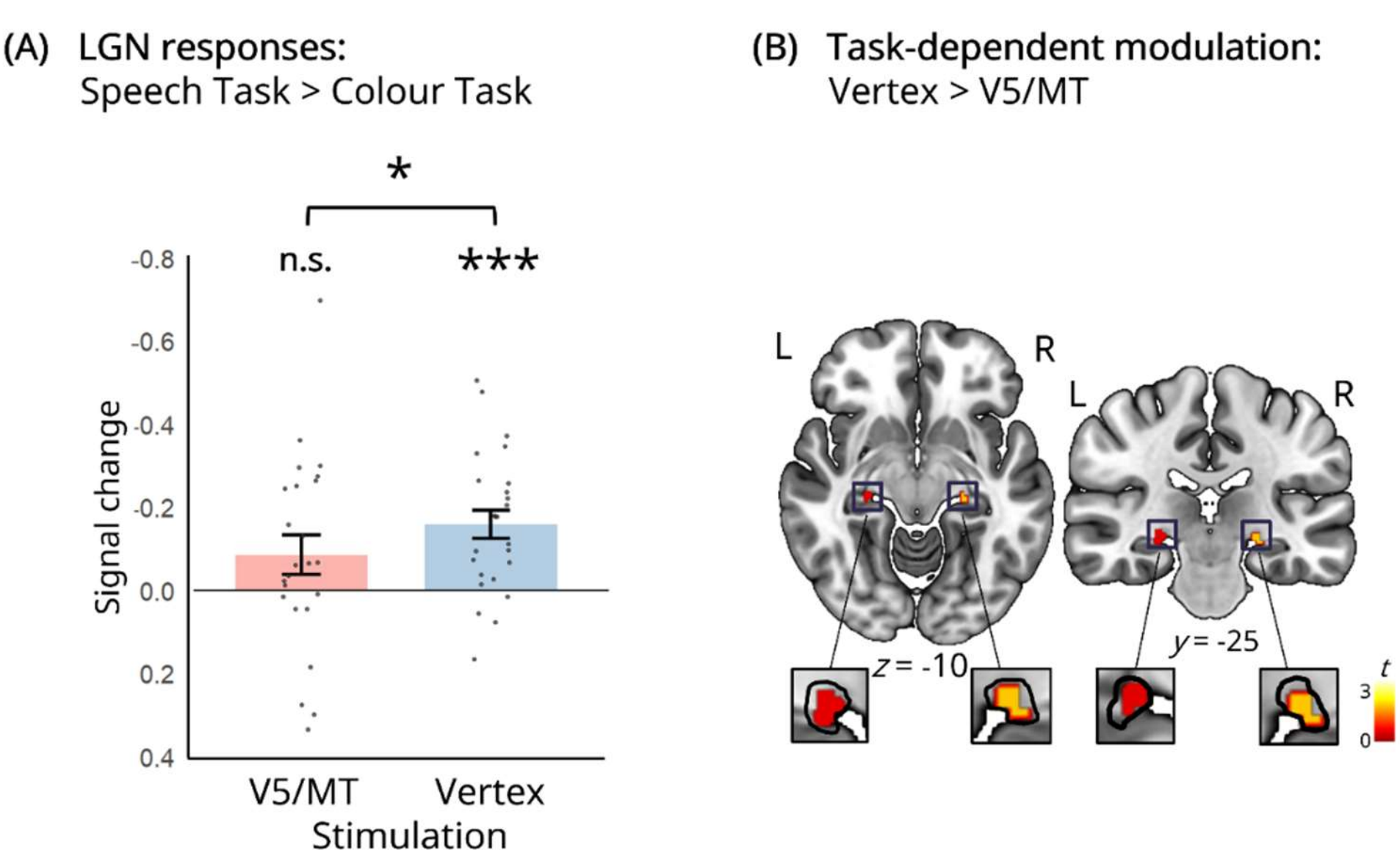

*Figure 3.* LGN responses. (A) Effects of Task on the signal change (in beta estimates) averaged across bilateral LGN (inverted axis). As expected, there was a significant task-dependent modulation following Vertex Stimulation, but not following V5/MT stimulation. Grey data points represent the individual means for each participant. Error bars represent 1 SEM. ***$p < 0.001$, *$p < 0.05$, n.s. = not significant. (B) Visualization of the interaction contrast of Task and Stimulation at the group level (estimated with SPM) in transverse and coronal views. The single-subject contrasts were MNI-normalised and masked with the ROI mask of the left and right LGN (indicated by the black outline).

**LGN-V5/MT functional connectivity**. We expected that V5/MT stimulation would reduce the task-dependent functional connectivity between V5/MT and LGN compared to control stimulation. For both PPI analyses (V5/MT as seed, LGN as seed), there was a significant task-dependent functional connectivity (V5/MT: $t(25) = 2.054$, $p = 0.001$ Bonferroni-Holm corrected, $d = 0.72$; LGN: ($t(25) = 2.251$, $p = 0.029$ Bonferroni-Holm corrected, $d = 0.48$) for Vertex stimulation, with a larger functional connectivity for the Colour Task compared to the Speech Task (mean difference V5/MT: 0.06; LGN: 0.02). In contrast, for V5/MT stimulation, analyses for both seeds did not show a significant task dependency (V5/MT: ($t(25) = 1.978$, $p = 0.091$ Bonferroni-Holm corrected, $d = 0.40$; LGN: $t(25) = 1.435$, $p = 0.158$ Bonferroni-Holm corrected, $d = 0.18$). The difference between stimulation conditions was not significant using V5/MT as seed ($t(25) = 1.155$, $p = 0.254$ Bonferroni-Holm corrected, $d = 0.22$; Figure 4), but was significant when using the LGN as seed ($t(25) = 2.455$, $p = 0.035$ Bonferroni-Holm corrected, $d = 0.41$, mean difference = 0.04; Figure 4).

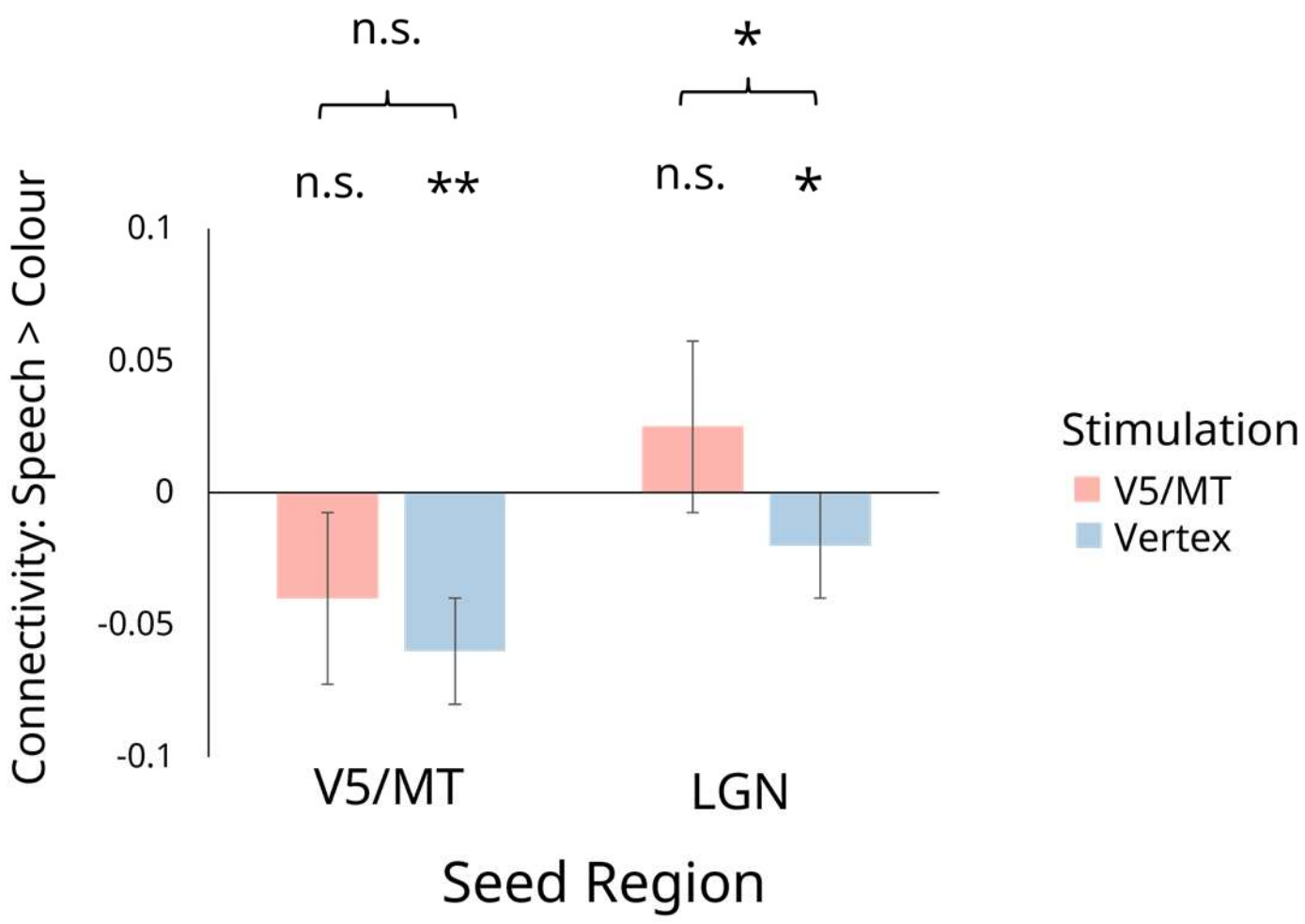

*Figure 4.* Task-dependent functional connectivity between V5/MT and LGN. The y-axis describes the mean beta estimate of the Task contrast. The left panel shows the results of the PPI analysis with V5/MT as seed region and LGN as target; The right panel shows the corresponding analysis using the LGN as seed and V5/MT as target region. Error bars represent 1 SEM. ** $p < 0.01$, * $p < 0.05$, n.s. = not significant.

**V5/MT responses.** As expected, we found a significant Task contrast with a larger BOLD response for the Speech compared to the Colour Task ($t(24) = 11.401$, $p < 0.001$, $d = 1.96$, Figure 5). The Task-Stimulation interaction was not significant ($t(24) = 0.320$, $p = 0.750$, $d = 0.04$).

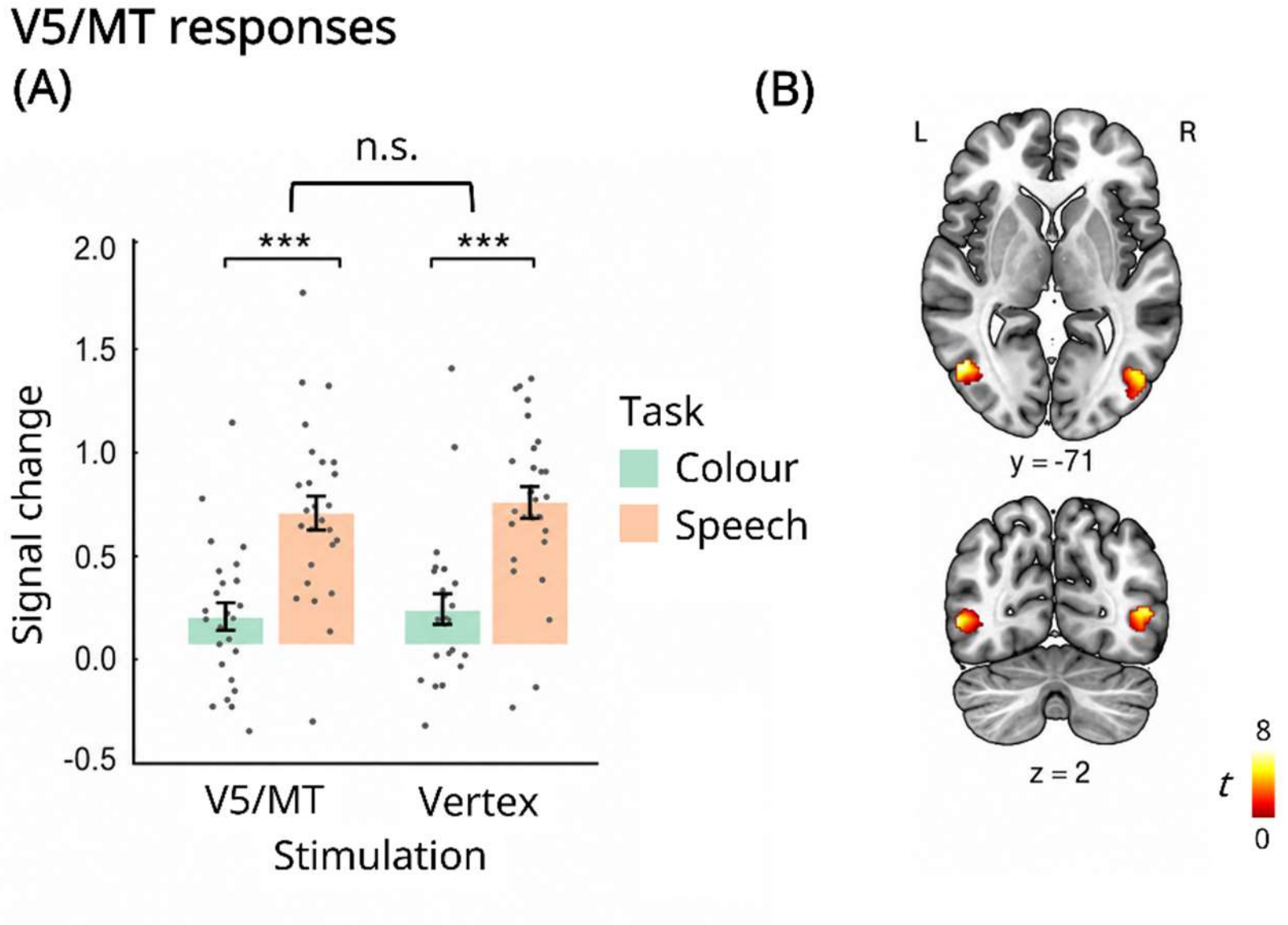

*Figure 5.* V5/MT responses. (A) Effects of Task (Colour/Speech) and Stimulation (V5/MT/Vertex) conditions on the signal change for bilateral area V5/MT. There was a stronger positive signal change of area V5/MT during the Speech Task compared to the Colour Task. Grey data points represent the individual means for each participant. Error bars represent 1 SEM. ***$p < 0.001$, n.s. = not significant. (B) For visualization purposes only, we normalized the single-subject contrast images to MNI space and masked the group contrast (estimated with SPM) with the ROI mask of V5/MT, shown in the transverse and coronal views.

**Behavioural results.** We hypothesized that inhibitory TMS over V5/MT would result in delayed RTs and larger RT ratios relative to Vertex stimulation, with larger effects for the Speech than the Colour Task. The linear mixed-effects model revealed a significant main effect of Stimulation for RT ratios ($\beta$ = -0.079; $t$ = -15.073; $p$ < 0.001; 95% *CI* [-0.089 0.069]; $d$ = 0.35; Figure 6A) and RTs: ($\beta$ = -61.63; $t$ = -9.24; $p$ = 0.002; 95% *CI* [-74.71 -48.56]; $d$ = 0.22): V5/MT stimulation led to larger RT ratios and slower response times than Vertex stimulation (Figure 6). We also found a significant Task-Stimulation interaction on RT ratios ($\beta$ = 0.064; $t$ = 9.002; $p$ < 0.001; 95% *CI* [0.050 0.078]; $d$ = 0.29) with an unexpected larger stimulation effect for the Colour Task (mean difference = 0.079; *SE* = 0.005; $p$ < 0.001) than for the Speech Task (mean difference = 0.016; *SE* = 0.005; $p$ = 0.014). Similarly, there was a significant Task-Stimulation interaction on RTs ($\beta$ = 53.34; $t$ = 5.89; $p$ < 0.001; 95% *CI* [35.59 71.08]; $d$ = 0.19; Figure 6B): the Stimulation effect was significant for the Colour Task (mean difference = 61.6; *SE* = 6.67; $p$ < 0.001), but not for the Speech Task (mean difference = 8.3; *SE* = 6.71; $p$ = 0.60). Lastly, there was a significant main effect of Task on RTs ($\beta$ = 838.88; $t$ = 10.17; $p$ = 0.001; 95% *CI* [782.76 895.00]; $d$ = 2.9). This was expected as solving the Speech Task required waiting for the critical viseme. Full model results for RTs and RT ratios, and analogous analyses on the average accuracies can be found as Supplementary Information.

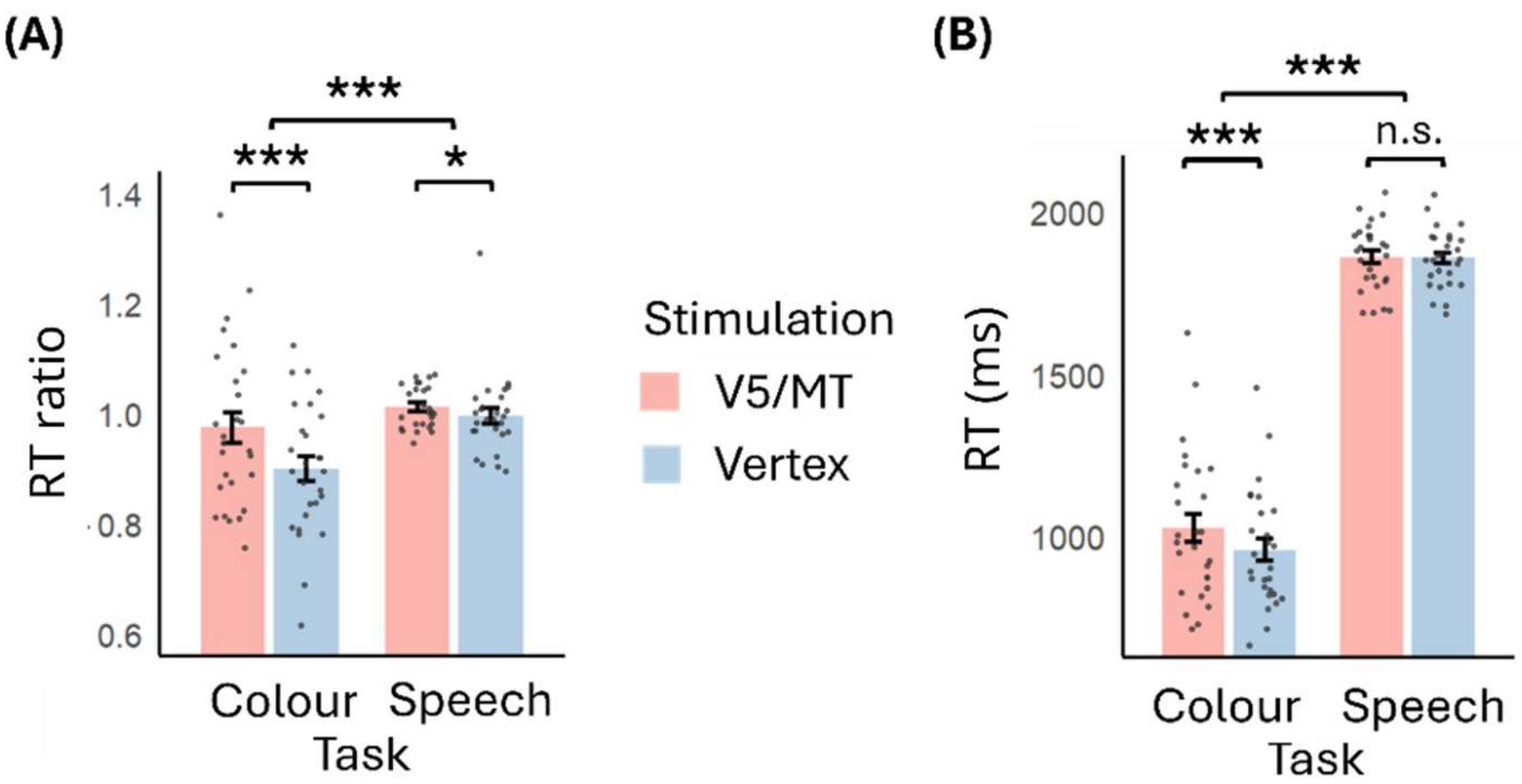

*Figure 6.* Response time ratios (RT ratio) and response times (RTs) during the fMRI experiment. (A) Bilateral stimulation over V5/MT increased RT ratios in both tasks compared with stimulation over the Vertex. Additionally, there was a significant interaction: the stimulation effect was larger for the Colour Task than for the Speech Task. (B) The RTs during fMRI showed a similar pattern as the RT ratios: V5/MT stimulation led to significantly slower RTs compared to Vertex stimulation, and there was a significant interaction with Task. The stimulation effect was only significant for the Colour Task. Grey data points represent the individual means for each participant. Error bars represent 1 SEM. ***$p$ < 0.001, **$p$ < 0.01, *$p$ < 0.05, n.s. = not significant.

**Correlation between neural responses and behaviour.** We examined whether LGN and V5/MT activity were associated with participants' visual speech recognition performance. First, we tested whether the Task-Stimulation interaction on RTs was positively correlated with the task-dependent modulation of LGN responses. We quantified the interaction analogous to the fMRI interaction contrast $\left(\left(RT_{V5MT,Speech} - RT_{Vertex,Speech}\right) - \left(\left(RT_{V5MT,Colour} - RT_{Vertex,Colour}\right)\right)\right.$. Task-dependent LGN modulation was estimated by extracting mean beta weights for each participant from the Task contrast. The correlation did not reach significance ($r(24) = 0.11$, $p$ = 0.579 Bonferroni-Holm corrected). We also tested whether the visual speech recognition performance was positively correlated with a) task-dependent LGN modulation, b) LGN responses during the Speech Task, and c) V5/MT responses during the Speech Task. Here we used mean accuracies as the measure of interest to compare the results to previous findings [7]. None of these correlations reached significance (see Supplementary Information).

## Discussion

We used TMS and fMRI to investigate whether response modulations in the human sensory thalamus are influenced by cerebral cortex function. The key finding was that task-dependent LGN modulation, i.e. significantly different LGN responses between a speech and a control task, was reduced after V5/MT stimulation compared to Vertex (control) stimulation. This reduction in task-dependent LGN modulation was accompanied by a corresponding reduction in task-dependent functional V5/MT-LGN connectivity. The results support our central hypothesis that feedback from V5/MT modulates LGN responses during task-relevant visual processing.

The findings align with the concept that cerebrocortical feedback serves to modulate the early processing of sensory input [9,42,43]. Animal studies demonstrated cerebral cortex influences on the sensory thalamus [8,9,11,13,44]. Previous human studies could only suggest that extensive feedback connections from the cerebral cortex to the thalamus are needed to adapt thalamic responses to specific tasks [6,7,45]. Our work provides direct causal evidence that such cortico-thalamic influence occurs in humans during active cognitive tasks. Potential mechanisms behind task-dependent thalamus modulation have been debated in previous research [9,46]. Our results identify cortical feedback as a key driver. One proposed account links thalamic response modulations to predictive coding models [47,48]. Corticothalamic feedback could contain predictive signals based on prior experience in order to fine-tune sensory processing in early sensory structures. In fact, subcortical sensory responses follow predictive coding principles in auditory processing [49-51]. It remains an open question whether these findings also extend to visual processing, including visual speech. Altogether, the results support the idea that V5/MT acts as key source of top-down feedback to the LGN, shaping early visual processing based on task demands.

The disruption caused by V5/MT stimulation on the task-dependent LGN modulation was accompanied by a delay of response times in the Speech Task. Moreover, V5/MT showed a larger BOLD response to the Speech Task than the Colour Task. This highlights the relevance of area V5/MT in the processing of visual speech cues. At the same time, our findings raise the question of how specialised V5/MT feedback to the LGN is to different cognitive tasks. For instance, we found a behavioural effect also on the Colour Task by inhibition of V5/MT, which is typically not considered colour-sensitive. Consistent with the behavioural results, functional V5/MT-LGN connectivity was stronger for the Colour Task. Two explanations could account for the behavioural findings. First, V5/MT stimulation might have affected both tasks in unanticipated ways. For instance, V5/MT can use colour information to define motion boundaries [52]. At the same time, TMS-induced changes in spatial suppression can enhance recognition performance in large moving patterns [53], which could have benefitted the Speech Task. Second, stimulation might have inhibited other colour-critical areas, such as V1 via V5/MT feedback [54] or deeper V5/MT-adjacent regions [55]. Furthermore, our findings on task-dependent functional connectivity suggests that the LGN-V5/MT pathway contributed to chromatic and contrast perception. Motion processing in V5/MT can be driven by luminance contrast [56–60] and anatomical studies have shown that koniocellular LGN layers provide S-cone input directly to V5/MT [19,61,62]. S-cone signals in this stream support processing of chromatic contrasts in natural illumination [63] and such inputs to V5/MT could hence support the detection of ecologically relevant features. Moreover, this pathway may be important for identifying visual changes [64]. For instance, participants could have relied on contrasts between speaker and background or subtle motion cues during the Colour Task, effectively creating fast-varying visual input to the LGN. In summary, signals between V5/MT and LGN may be flexibly recruited by tasks in which colour aspects provide cues about rapidly changing images. Our results imply that V5/MT feedback may generally support processing of dynamic stimuli, even when tasks are not motion or speech-specific.

Similarly, our results suggest that LGN modulation does not directly translate to visual-only speech recognition behaviour. Visual speech recognition accuracy and LGN responses were not correlated, unlike in previous work, potentially due to methodological differences such as intra- versus inter-speaker speech recognition [7], but also possibly due to limited statistical power. While task-dependent modulation of the medial geniculate body (MGB), i.e. sensory auditory thalamus, correlates with auditory speech recognition performance [6,65], this has not been observed for the LGN and visual speech [7]. Such discrepancy can reflect fundamental differences between LGN-mediated visual speech recognition versus MGB-mediated auditory speech recognition: visual speech is rarely processed in isolation but instead aids auditory speech recognition, particularly under noise [2,3,66], where also MGB modulation is pronounced [5]. It therefore remains an open question how the LGN relates to speech recognition behaviour in audiovisual settings.

While our results provide strong evidence for corticothalamic feedback during visual processing in humans, the observed thalamic responses likely arise from multiple interacting processes and cannot be unequivocally attributed to top-down modulation alone. For instance, in contrast to previous studies demonstrating task-dependent modulations of sensory thalamic nuclei during speech recognition [5-7], we observed negative LGN BOLD signal changes overall, with a more negative signal change for the Speech Task. We can only speculate on the reasons for this discrepancy - for instance methodological differences such as the use of a colour task as control condition. Furthermore, LGN responses could have been driven by additional factors or unanticipated response mechanisms, e.g. attentional effects or figure-ground modulation [67–72]. The interpretation of negative signal changes remains unclear: while they may simply reflect a deviation from baseline or neural suppression [73–78], baseline LGN responses can also be elevated, for instance in anticipation of visual stimulation [69], resulting in a relative suppression during the task. Regardless the nature of sensory processes, our results show that the adaptation of LGN responses according to task demands depends on the integrity of V5/MT. This influence was corroborated by the effects of V5/MT stimulation on task-dependent functional connectivity between V5/MT and LGN. However, this effect was only significant using the LGN as seed, but not V5/MT. PPI analyses can be sensitive to differences in signal amplitude and signal-to-noise ratio between seeds [79,80]. Consequently, it remains speculative whether top-down influences differ in magnitude between cortical and subcortical regions. Despite these limitations, the findings support the notion that V5/MT may serve as a key source of top-down feedback to the LGN.

The results extend functional [20] and structural [21,23] human evidence on direct V5/MT-LGN connections, assumed to be critical for processing high-frequency visual information [15,16,19]. Sensory thalamic nuclei, V5/MT, and their connections are particularly relevant for neurodevelopmental disorders affecting human communication, such as autism-spectrum disorder and developmental dyslexia [23,81]. The LGN is central to the sensory processing in such conditions [23,82,83], and understanding its mechanisms and functional relevance therefore provides a basis for research into the neural basis of communicative deficiencies. In conclusion, our findings provide first causal evidence that feedback from V5/MT contributes to the task-dependent modulation of the LGN during active visual processing.

## Acknowledgements

The study was funded by the ERC-Consolidator Grant SENSOCOM 647051, the Eranet-neuron grant ReDyslexia (Federal Ministry of Education and Research/Bundesministerium für Bildung und Forschung, grant agreement ID 01EW2213), and the Saxon State Ministry of Science, Culture and Tourism (SMWK). Special thanks to Kira Eckert, Livia Rühr, and Isabel Eschelor for their help with organizing and conducting the experiment.

# Supplementary Information

## Methods

### Participant exclusion criteria

None of the participants reported any contraindications for MRI and TMS, or a history of psychiatric or neurological disorders. Ten participants were excluded from analysis: five did not complete the study, three performed below 60% accuracy, and two for an insufficient quality of their fMRI data.

All participants were screened for conditions associated with altered visual speech recognition, including autism spectrum disorder [S1] and developmental dyslexia [S2]. Screening comprised a questionnaire about dyslexia diagnoses and the German version of the autism-spectrum quotient [S3]. No participant reported a diagnosis of developmental dyslexia or scored below the clinical autism-spectrum quotient cutoff of 26 points ($M$ = 16.15, $SD$ = 5.38). Colour vision was assessed with six Ishihara plates (mean accuracy = 96.83%, $SD$ = 0.07% [S4]). All participants demonstrated normal or corrected-to-normal vision (decimal visual acuity > 0.9) on the Freiburg Visual Acuity Test [S5].

### Experimental Design - Visual Speech and Colour Recognition Task

**Stimulus randomisation.** Each block contained three syllables, three colours, and one speaker, with all syllables starting with the same viseme. Colours within a block were separated by two or three steps in the colour spectrum.

**Eye tracking pilot.** Eye movements and fixations can modulate LGN responses [S6]. To ensure that both tasks would be comparable in terms of eye movement patterns, we collected eye-tracking data from three participants not included in the main sample prior to the data acquisition for the main experiment. Participants performed one session of the experimental tasks while their eye movements were monitored via an eye tracking system (EyeLink 1000 Plus, SR-Research, Canada). The data was preprocessed with Eyelink Data Viewer (SR Research Ltd., Canada) and analysed with R (version 3.6.3 [S7]). We assessed the average number and duration of fixations on the presented videos and excluded any fixations that occurred after the participant's button press. The fixations were categorised into four rectangular areas of interest (AOI): the speaker's eyes, nose, mouth, and a fourth category for all fixations outside these regions. For the statistical analysis, we carried out two repeated-measure ANOVAs with a) the average number of fixations per trial and b) the average duration of fixations as dependent variables. The within-subject factors were Task (Speech vs Colour) and AOI (Eyes, Nose, Mouth, and Periphery). Importantly for the experimental design, there was no significant interaction of Task and AOI on the average fixation duration ($F(1,3) = 0.64$; $p = 0.599$; $\eta^2 = 0.09$) or number of fixations ($F(1,3) = 0.21$; $p = 0.888$; $\eta^2 = 0.02$). There were also no significant main effect of Task on the average fixation duration ($F(1, 3) = 1.89$; $p = 0.188$; $\eta^2 = 0.09$) and number of fixations ($F(1,3) = 1.83$; $p = 0.195$; $\eta^2 = 0.07$), i.e. participants did not show different eye movements between tasks. In addition, the main effects of AOI on the average number of fixations ($F(1,3) = 3.16$; $p = 0.054$; $\eta^2 = 0.34$), and average duration of fixations ($F(1,3) = 0.56$; $p = 0.653$; $\eta^2 = 0.08$) were not significant. In this case, the number of fixations between AOIs was close to significance, meaning that there was a trend for a larger number of fixations on the eye region than the mouth region, independent of the task. The lack of differences in eye movements between tasks in such experimental design agrees with previous research [S8]

## Follow-Up Questionnaires

At the end of the final session, participants completed a questionnaire to report perceived differences between the two TMS conditions, any TMS-related side effects, and the strategies used in the Speech and Colour Tasks.

| Perceived differences between sessions | *n(%)* |
|---|---|
| *None* | 9 (35%) |
| *Difference in intensity/effect of the stimulation* | 13 (50%) |
| Thereof: | |
| *V5/MT stimulation more intense* | 10 (38%) |
| *Vertex stimulation more intense* | 3 (12%) |
| *Change in side effects* | 3 (12%) |
| *Difference in duration* | 1 (4%) |

***Supplementary Table 1.*** Did you perceive any differences between the two TMS sessions? If yes, which? (German: “Haben Sie Unterschiede zwischen den beiden TMS-Sitzungen wahrgenommen? Falls ja, welche?“)

| Perceived side effects | *n(%)* |
|---|---|
| *None* | 20 (77%) |
| *Mild headache* | 5 (19%) |
| *Fatigue* | 1 (4%) |

***Supplementary Table 2.*** Did you experience any side effects during or after the stimulation? If yes, which ones? (German: “Sind wa hrend oder nach der Stimulation Nebenwirkungen aufgetreten? Falls ja, welche?“)

| Used strategies | *n(%)* |
|---|---|
| Colour Task: | |
| *Focus on facial features* | 11 (42%) |
| Thereof: | |
| *Forehead* | 8 (31%) |
| *Edges* | 2 (8%) |
| *Not specified* | 1 (4%) |
| *Holistic perception* | 4 (15%) |
| Speech Task: | |
| *Internal speaking* | 8 (31%) |
| *Imagining words* | 7 (27%) |
| *Focus on mouth* | 6 (23%) |
| *Focus outside of mouth* | 1 (4%) |

***Supplementary Table 3.*** Did you use any strategies for absolving the speech and/or colour task? If yes, which ones? (German: “Haben Sie irgendwelche Strategien zum Absolvieren der Silben- und/oder Farbe-Aufgabe angewandt? Falls ja, welche?”)

## Transcranial magnetic stimulation

**Active motor threshold measurement.** Single TMS pulses were delivered to the left primary motor cortex (M1), starting at 50% stimulator output, while participants were asked to slightly tense their right index finger muscle by pressing it against the thumb. Stimulation intensity was reduced until 5 of 10 pulses elicited motor-evoked potentials exceeding 50 µV. Motor thresholds are stable across days [S9-11]. M1 target coordinates were derived from a standard site ($x$ = 37, $y$ = −21, $z$ = 58 [S12]) and transformed into individual native space.

**Neuronavigation.** We used stereotactic neuronavigation (BrainSight, Rogue Research, Canada) to precisely position the TMS coil over each participant's individual V5/MT, vertex, and M1 coordinates. The coil was set over the system-indicated entry points corresponding to the respective coordinates. These entry points were those sites on the participant's scalp with the shortest distance to the target coordinates. We counterbalanced whether participants received stimulation over the anterior or posterior Vertex first, as well as whether left or right V5/MT was stimulated first. For the Vertex area, two sites located 1.5 cm anterior and posterior to the computed vertex coordinate were determined. The vertex coordinates were transformed from MNI space into each participant's individual space. These coordinates correspond to the location of the Cz electrode in the 10–20 EEG system (MNI coordinates: $x = 0.8$, $y = -14.7$, $z = 73.9$) projected onto the MNI cortical surface [S13]. For left V5/MT (mean MNI coordinate: $x = -45.04 \pm 3.69$, $y = -69.85 \pm 4.58$, $z = 2.22 \pm 3.89$) and right V5/MT (mean MNI coordinate: $x = 45.00 \pm 3.71$, $y = -73.19 \pm 3.98$, $z = 4.11 \pm 4.31$), the coordinates were retrieved from each participant's V5/MT functional localiser. We computed the average induced electric fields of the stimulation targets (Supplementary Figure 1).

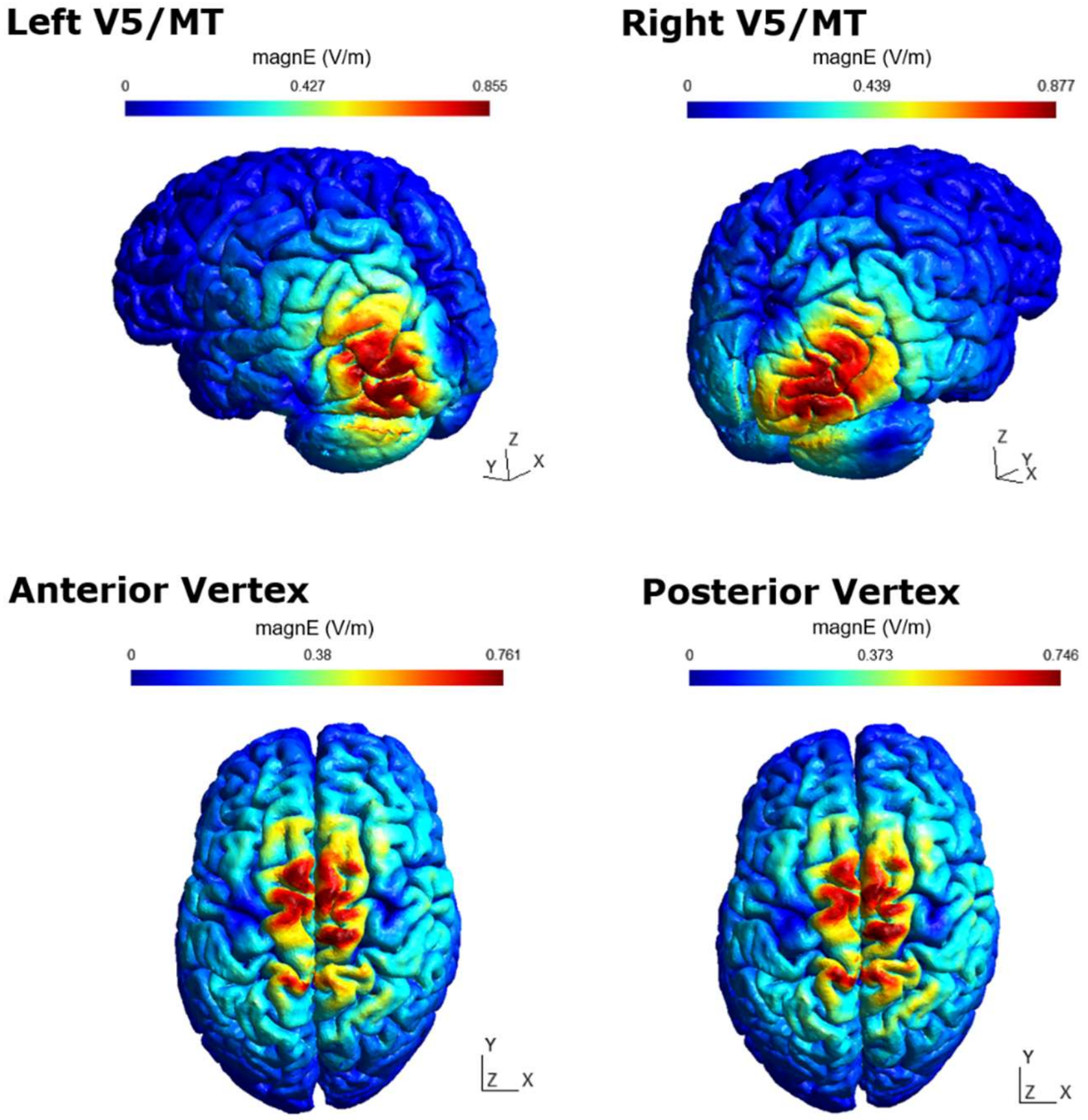

***Supplementary Figure 1.*** The average magnitude of the induced electric field (magnE) across participants displayed on the normalised cortical surface of an example subject ("Ernie" head model) for the four stimulation sites: left V5/MT, right V5/MT, anterior Vertex region, and posterior Vertex region, computed with SimNIBS software (version 4.1.0 [S14]).

## Functional localisation of V5/MT

We used a well-established functional localiser to identify V5/MT in each individual participant by running a conventional "moving versus static" paradigm [S15,16]. The participants viewed random-dot kinematograms (RDKs) that consisted of either a single frame of static dots (static condition) or dots moving coherently inward or outward (moving condition). The dots were white and 0.1° in diameter, surrounding a grey fixation point of 0.2° diameter located at the centre of the screen against a black background. In the moving condition, a total of 250 dots were generated, moving at a speed of 4.7°/s. The stimuli were created using MATLAB R2019b (version 9.6.0.1047502) and Psychtoolbox (version 3.0.15; [S17–19]). The paradigm comprised 32 blocks, randomly alternating between static and moving conditions (8× inward moving, 8× outward moving, 16× static).

### Functional localisation of the LGN

To identify the LGN, we used a functional localiser [S20]. The stimuli consisted of flickering radial checkerboards with 100% contrast and a contrast polarity reversal at 4 Hz. Alternatingly, the left (or right) half of the screen was covered with the checkerboard pattern, while the other half consisted of a grey background colour. Each of the 16 blocks lasted 16 s, resulting in a total run time of 5 minutes. Participants were instructed to focus on a white fixation at the centre of the screen.

## Results

### Behavioural responses

We analysed behavioural measures on a trial-by-trial basis by creating linear mixed effects models. The random-effects structure of the mixed-effects models was determined using a backward model selection, in which random-effects terms that accounted for the least variance were removed one by one until the fitted mixed model converged.

| | **Response time ratio (fMRI/baseline)** | | | | |
|---|---|---|---|---|---|
| **Predictors** | | | | | |
| | *β* | *CI* | *t* | *SE* | *p* |
| Intercept | 0.977 | 0.941, 1.014 | 52.636 | 0.019 | **< 0.001** |
| Stimulation | -0.079 | -0.089, -0.069 | -15.073 | 0.005 | **< 0.001** |
| Task | 0.034 | 0.001, 0.069 | 1.988 | 0.016 | 0.102 |
| Stimulation * Task | 0.064 | 0.050, 0.078 | 9.002 | 0.064 | **< 0.001** |
| **Random Effects** | | | | | |
| | *Variance* | *SD* | | | |
| Subject | | | | | |
| Intercept | 0.009 | 0.093 | | | |
| Stimulation | 0.007 | 0.086 | | | |
| Stimulus | | | | | |
| Intercept | 0.006 | 0.074 | | | |
| Stimulation | 0.001 | 0.036 | | | |
| Task | 0.005 | 0.070 | | | |

***Supplementary Table 4.*** Linear mixed-effects model of Stimulation and Task on pre-stimulation (baseline) and post-stimulation (fMRI) response time ratios. Significant p-values are marked in bold. The final model for the response time ratios comprised a random intercept by Subject and a random slope by Subject for Task, and a random intercept by Stimulus and a random slope by Stimulus for Stimulation and Task.

| Predictors | Response time (fMRI) | | | | |
|---|---|---|---|---|---|
| | *б* | *CI* | *t* | *SE* | *p* |
| Intercept | 1025.79 | 953.14, 1094.43 | 28.40 | 36.04 | **< 0.001** |
| Stimulation | -61.63 | -74.71, -48.56 | -9.24 | 6.67 | **0.002** |
| Task | 838.88 | 782.76, 895.00 | 29.30 | 28.63 | **< 0.001** |
| Stimulation * Task | 53.34 | 35.59, 71.08 | 5.89 | 9.05 | **< 0.001** |
| **Random Effects** | | | | | |
| | *Variance* | *SD* | | | |
| Subject | | | | | |
| Intercept | 34189 | 185 | | | |
| Stimulation | 20333 | 143 | | | |
| Stimulus | | | | | |
| Intercept | 7536 | 87 | | | |
| Stimulation | 2463 | 50 | | | |
| Task | 15754 | 126 | | | |

***Supplementary Table 5.*** Linear mixed-effects model of Stimulation and Task on response times (in milliseconds). Significant p-values are marked in bold. The final model for the response times comprised a random intercept by Subject and a random slope by Subject for Task, and a random intercept by Stimulus and a random slope by Stimulus for Stimulation and Task.

| Predictors | Accuracy Ratio (fMRI/baseline) | | | | |
|---|---|---|---|---|---|
| | *б* | *CI* | *t* | *SE* | *p* |
| Intercept | 1.003 | 0.967, 1.040 | 53.974 | 0.019 | **< 0.001** |
| Stimulation | 0.033 | -0.025, 0.092 | 1.119 | 0.030 | 0.272 |
| Task | 0.065 | 0.005, 0.125 | 2.137 | 0.030 | **0.040** |
| Stimulation * Task | 0.025 | -0.002, 0.052 | 1.780 | 0.014 | 0.075 |
| **Random Effects** | | | | | |
| | *Variance* | *SD* | | | |
| Subject | | | | | |
| Intercept | 0.008 | 0.088 | | | |
| Stimulation | 0.022 | 0.147 | | | |
| Task | 0.022 | 0.148 | | | |
| Stimulus | | | | | |
| Intercept | 0.006 | 0.079 | | | |
| Task | 0.013 | 0.112 | | | |

***Supplementary Table 6.*** Linear mixed-effects model of Stimulation and Task on the pre-stimulation (baseline) and post-stimulation (fMRI) accuracy ratios. Significant p-values (Bonferroni-Holm corrected) are marked in bold. The final model for accuracy ratio included a random intercept by Subject and a random slope by Subject for Task and Stimulation, and a random intercept by Stimulus and a random slope by Stimulus for Task.

| Predictors | Accuracy (fMRI) | | | | |
|---|---|---|---|---|---|
| | *β* | *CI* | *t* | *SE* | *p* |
| Intercept | 0.804 | 0.784, 0.823 | 81.334 | 0.010 | **< 0.001** |
| Stimulation | 0.024 | 0.010, 0.039 | 3.291 | 0.007 | **0.002** |
| Task | 0.027 | 0.002, 0.051 | 2.156 | 0.012 | 0.072 |
| Stimulation * Task | -0.009 | -0.029, 0.012 | -0.893 | 0.010 | 0.804 |
| **Random Effects** | | | | | |
| | *Variance* | *SD* | | | |
| Subject | | | | | |
| Intercept | 0.002 | 0.041 | | | |
| Task | 0.002 | 0.048 | | | |
| Stimulus | | | | | |
| Intercept | 0.005 | 0.069 | | | |
| Task | 0.008 | 0.089 | | | |

***Supplementary Table 7.*** Linear mixed-effects model of Stimulation and Task on the mean accuracies. Significant p-values (Bonferroni-Holm corrected) are marked in bold. The final model for the accuracy during fMRI only comprised a random intercept by Subject and a random slope by Subject for Task, and a random intercept by Stimulus and a random slope by Stimulus for Task.

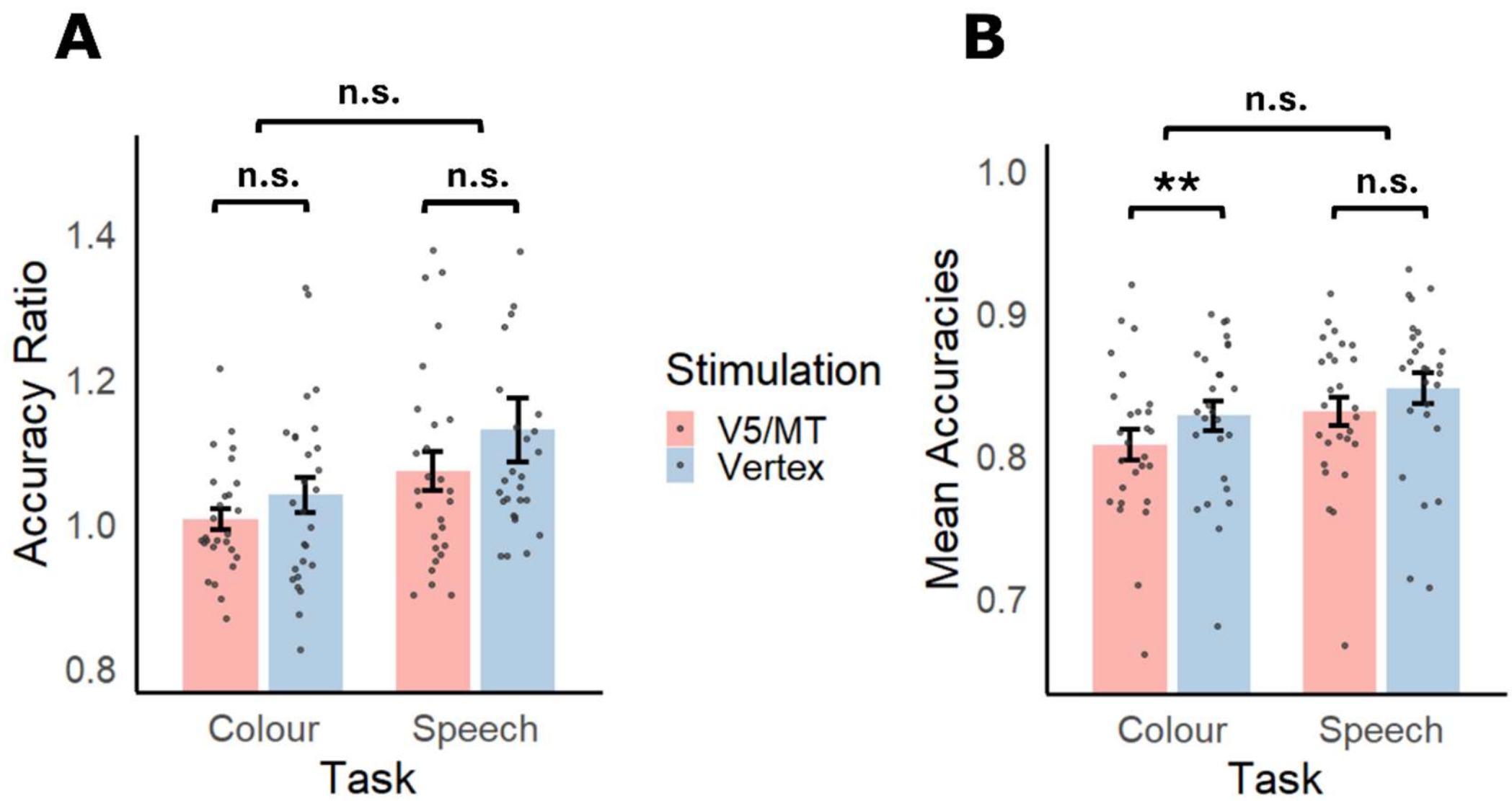

***Supplementary Figure 2.*** Effects of inhibitory TMS (continuous theta burst stimulation) on the accuracy ratios and mean accuracies in the Speech and Colour Task. A) There was no significant effect of Stimulation for either task. Additionally, there was a significant effect of Task: the Speech Task showed a larger practice effect between baseline and fMRI compared to the Colour Task. There was no significant interaction of Task and Stimulation. B) The accuracies during fMRI alone showed a significant main effect of Stimulation: V5/MT decreased task accuracies compared to Vertex stimulation. However, after correction for multiple comparisons, this effect was significant only for the Colour Task. There was no significant interaction of Task and Stimulation. Grey data points represent the individual means for each participant. Error bars represent 1 SEM. **p < 0.01, n.s. = not significant.

**Interaction of neural responses and behaviour**

| Correlation of mean accuracy (% correct) to signal change | | |
|---|---|---|
| | *r (df = 24)* | *p* |
| **LGN: Speech > Colour** | 0.001 | 0.996 |
| Vertex stimulation | -0.04 | 2.453 |
| V5/MT stimulation | -0.03 | 1.747 |
| **LGN: Speech** | -0.24 | 0.726 |
| Vertex stimulation | -0.20 | 0.664 |
| V5/MT stimulation | -0.10 | 0.640 |
| **V5/MT: Speech** | -0.18 | 0.764 |
| Vertex stimulation | -0.06 | 0.787 |
| V5/MT stimulation | -0.23 | 0.795 |

***Supplementary Table 8***. Pearson's product-moment correlations between the mean accuracies of the Speech Task during fMRI and the signal change by Task in the LGN, the LGN response during the Speech Task, and the V5/MT response during the Speech Task. We additionally estimated the correlations for sessions following V5/MT stimulation and Vertex stimulation, separately. The additional correlations were Bonferroni-Holm corrected for multiple comparisons. None of the correlations were significant.

**Supplementary References**